**Role of dipolar interactions in a system of Ni nanoparticles studied by magnetic susceptibility measurements**

S. H. Masunaga and R. F. Jardim*

Instituto de Física, Universidade de São Paulo, CP 66318, 05315-970 São Paulo, SP, Brazil

P. F. P. Fichtner

Escola de Engenharia, UFRGS, 91501-970 Porto Alegre, RS, Brazil

J. Rivas

Departamento de Física Aplicada, Universidade de Santiago de Compostela, Campus Universitario, 15706 Santiago de Compostela, Spain

The role of dipolar interactions among Ni nanoparticles (NP) embedded in an amorphous $SiO_2$/C matrix with different concentrations has been studied performing ac magnetic susceptibility $\chi_{ac}$ measurements. For very diluted samples, with Ni concentrations < 4wt % Ni or very weak dipolar interactions, the data are well described by the Néel-Arrhenius law. Increasing Ni concentration to values up to 12.8 wt % Ni results in changes in the Néel-Arrhenius behavior, the dipolar interactions become important, and need to be considered to describe the magnetic response of the NPs system. We have found no evidence of a spin-glasslike behavior in our Ni NP systems even when dipolar interactions are clearly present.

## I. INTRODUCTION

A particle of a ferromagnetic (FM) material is believed to be more stable when consisting of only one magnetic domain, provided that its size is below a certain critical value.[1] The critical radius $R_C$ of ordinary FM nanoparticles (NPs) have been estimated to be in the range of 10–1000 nm,[2] a value calculated to be ~ 30 nm for nickel Ni.[3] Ni is a prototype of FM material but when it is reduced below its critical size $R_C$, thermal agitation will lead to a superparamagnetic (SPM) state instead of stable magnetization. The reversion of the magnetic moment of an isolated single-domain particle with volume $V$ and anisotropy constant $K$ over a uniaxial anisotropy energy barrier $E_a$ is characterized by the Néel-Arrhenius relaxation time,[4]

$$\tau^N = \tau_0^N \exp\left(\frac{E_a^N}{k_B T}\right). \tag{1}$$

where $\tau_0^N$ in the range of $10^{-12} - 10^{-9}$ s is a characteristic attempt time for FM NPs, $k_B$ is the Boltzmann constant, and $E_a^N = KV$. In a system comprised of fine particles, with a distribution of particle size and various sources of anisotropy, $V$ is replaced by a mean value $\langle V \rangle$ and the activation energy is then related to an effective anisotropy constant $K_{eff}$. For observation times $\tau_{obs}$ much longer than the relaxation time $\tau$( $\tau_{obs} \gg \tau$), the magnetic moments $\vec{\mu}$ of the NPs reach the thermal equilibrium, characterizing the SPM behavior. On the other hand, for $\tau_{obs} \ll \tau$, the $\vec{\mu}$ are blocked by the energy barrier, a state corresponding to a stable magnetization below a certain blocking temperature $T_B$, which is defined for $\tau_{obs} = \tau$.

* rjardim@if.usp.br

The dipole-dipole interaction is always present in systems comprised of magnetic NPs and can be only considered the most relevant interaction provided that the agglomeration of NPs is absent and that they are stabilized within an isolated matrix or separated by a surfactant.[5] Due to the ~ $10^3$ $\mu_B$ order of the magnetic moment of the NPs, the magnitude of the dipolar interaction becomes appreciable so the manifestation of its effect is easily observed in ordinary magnetic characterizations.[5] For lightly diluted samples, the noninteracting case is treated as a good approximation. However, even for very diluted samples, the interaction strength can reach units or tens of kelvins, becoming a relevant part of the total energy of the system. A rough estimate of the temperature $T_0^{Eq}$ related to the dipolar interaction strength between two NPs is given by the relationship[6]

$$T_0^{Eq} = \frac{\langle \mu \rangle^2}{k_B r^3} . = \frac{M_S^2 \langle V \rangle}{k_B} x_V, \tag{2}$$

where $\langle \mu \rangle$ is the mean magnetic moment of a log-normal distributed system, $r$ is the average interparticle distance assuming a regular arrangement of particles in the sample, $x_V$ is the volume concentration of particles, $\langle V \rangle$ is the mean particle volume, $M_S$ is the saturation magnetization, and $k_B$ is the Boltzmann constant.

Equation (1) has been used frequently and accounts for the magnetic properties of ensembles of noninteracting particle systems.[7] However, when the concentration of NPs increases, the energy barrier is modified by the magnetic interaction between granules. Under this circumstance, the Vogel-Fulcher law,

$$\tau^V = \tau_0^V \exp\left(\frac{E_a^V}{k_B(T - T_0^V)}\right). \tag{3}$$

where $T_0^V$ is the strength of interparticle dipolar interaction, is useful to describe the magnetic properties of systems with weak magnetic interactions, provided that $T_B \gg T_0^V$.[8] The same approach is also used for describing the magnetic properties of spin-glass systems, where $T_0$ is now understood as the ideal glass temperature for real glasses.[9] Although the Vogel-Fulcher approach has been used for different systems, the frequency $f = 1/\tau$ sensitivity between the blocking temperature $T_B(f)$ in SPM systems and the freezing temperature $T_f(f)$ in spin glasses is quite different in nature.[10]

Separating the contribution of the dipolar interaction in systems with increasing particle concentration constitutes a difficult task. This is mainly due to the intricate details in producing samples with different concentrations of NPs without changing their size distributions appreciably. Increasing the particle size usually lead to an increase in the dipolar interaction while different sizes introduce size effects.[10]

We discuss here the ac magnetic susceptibility $\chi_{ac}$ data at zero applied dc magnetic field of five samples with different concentrations of Ni NPs. Even though we have already studied systems similar to the ones discussed here, roughly determining a dipolar energy added to the barrier energy by comparing only two samples with different Ni concentrations and average radius,[11,12] the sample preparation method has been improved and we have synthesized Ni NPs in a larger range of concentration. Furthermore, we discuss the results of $\chi_{ac}$ in samples with similar radius $R$ ~2.5 nm $\ll R_C$ and width of particle distribution, providing a systematic study of the magnetic dynamics of Ni NPs. Mostly due to the very small ac applied magnetic field used in all measurements, $H_{ex}$ = 1 Oe, only a slight modification is caused to the barrier energy. Therefore the relaxation time $\tau$ in in the absence of magnetic field has been used as a good approximation.[10] The combination of these characteristics has allowed us to systematic study the effect of dipolar interaction on the magnetic properties of these Ni NPs. Moreover, by studying the frequency sensitivity of the peak temperature in $\chi_{ac}$ curves we were able to assure that spin-glasslike behavior is absent in our Ni NPs systems.

## II. EXPERIMENTAL PROCEDURE

Ni nanoparticles embedded in an amorphous matrix, which is composed of $SiO_2$ and C ($SiO_2$ /C), were prepared through a method in which silicon oxianions and the metal cation (Ni) are immobilized within a polymeric matrix based on polyester. Initially, citric acid was dissolved in ethanol and tetraethyl orthosilicate [$Si(OC_2H_5)_4$, 98%]. Nickel (II) nitrate hexahydrate [$Ni(NO_3)_2{\cdot}6H_2O$, 99.9985%] was added to the latter mixture and mixed for homogenization. Ethylene glycol was thus added to promote the polymerizing reaction when the solution is heated up to 80 °C. The resulted solid resin was heat treated at 300 °C for 3 h, ground in ball mill and then pyrolyzed in $N_2$ atmosphere at 500 °C. The pyrolysis of the precursor resulted in a $CO/CO_2$ rich atmosphere which together with the $N_2$ flux, promote the reduction in the Ni citrate. Further details of method employed for producing the samples are described elsewhere.[13] The Ni content of the samples ($x$ =1.9, 2.7, 4.0, 7.9, and 12.8 wt % Ni) was determined by inductively coupled plasma atomic emission spectrometry (ICP-AES).

X-ray powder-diffraction (XRD) patterns were taken using a D8 Advance Bruker-ASX diffractometer with Cu$K\alpha$ radiation. A 2θ angular scan between 20$^{o}$ and 80$^{o}$ was made in all samples with 2θ accuracy of ~0.05$^{o}$ at room temperature. The XRD diagrams indicated that all samples are single phase and comprised of crystalline fcc Ni and an amorphous phase related to the $SiO_2$/C matrix.

The size distribution and the morphology of the Ni NPs were studied by transmission electron microscopy (TEM) on two microscopes: (i) JEOL JEM-2010 operating at 200 kV and (ii) Philips CM-12, operating at 120 kV. Measurements of the ac magnetic susceptibility $\chi_{ac} = \chi' + i\chi''$ were performed in powder samples by using two Quantum Design apparatuses: a superconducting quantum interference device magnetometer and a physical property measurement system. The measurements were performed under zero dc magnetic field, an excitation field of 1 Oe, and driving frequencies varying over five decades, or more appropriately, from 0.033 to 9999 Hz.

## III. RESULTS AND DISCUSSION

All the five samples containing Ni concentrations $x$=1.9, 2.7, 4.0, 7.9, and 12.8 wt % Ni were synthesized by the above-mentioned method. TEM analyses were performed in the series to confirm the presence of Ni NPs and to determine their size distribution. Typical images from the TEM investigation for the samples studied, accompanied by their histograms, are displayed in Fig. 1. The histograms were constructed by counting over 1400 Ni NPs spread out on different regions of the TEM grid and considering a log-normal distribution. The images from TEM investigation indicate that the Ni NPs are homogeneously distributed within the amorphous $SiO_2$/C matrix, with no sign of agglomeration, and have nearly spherical shape. For example, the relevant size parameters of the sample with 4.0 wt % Ni were a mean diameter of $\langle d \rangle$ = 5.0 nm, a median diameter $d_0$ =4.9 nm, and a size distribution width of $\sigma_d$ = 0.24, as estimated by the log-normal distribution function. In addition to this, as displayed in Table I, all samples have quite similar average diameter and size-distribution width. The average distance $r$ between NPs is believed to decrease with increasing Ni concentration therefore increasing the coupling between particle magnetic moments. We have found that $r$, estimated from the Ni concentration in the samples and in excellent agreement with the ones extracted from the TEM images, roughly varies from 14 nm for the most concentrated sample (12.8%) to 21 nm for the most diluted one (1.9%), as displayed in Table I. The latter value of $r$ = 21 nm is essentially the same when compared to the one of the sample with 2.7 wt % Ni due to the small difference in the average radius of the samples. In any event, these features are important for studying the role of the dipolar interactions in this series, as discussed below.

The temperature dependence of both the real $\chi'$ and imaginary $\chi''$components of the ac magnetic susceptibility for two selected samples, 1.9 and 12.8 wt % Ni or more appropriately the most diluted and the most concentrated one, respectively, are shown in Fig. 2. Some features of interest in these curves are: (i) the occurrence of a frequency-dependence rounded maximum at $T'_m$ and $T''_m$, on both $\chi'$ and $\chi''$ components; (ii) increasing $f$ results in a shift of both $T'_m$ and $T''_m$, to higher temperatures; (iii) $\chi'$ is frequency independent at high temperatures $T \gg T'_m$, indicating a SPM behavior of the Ni NPs (Refs. 10 and 14); (iv) a frequency-dependent behavior of $\chi'$ for $T \leq T'_m$ signaling the blocking process of the Ni NPs.[10] We notice that at very low frequency (~0.033 Hz), the $\chi'$ component of $\chi_{ac}$ peaks at $T'_m$ ~ 12 K and $T'_m$ ~ 36 K for samples with 1.9 and 12.8 wt % Ni, respectively. We have observed a similar trend in ac magnetic susceptibility curves for other samples, that exhibited peaks at $T_m^{'}$ ~ 15, 16, and 23 K, for Ni content of 2.7, 4.0, and 7.9 wt %, respectively. We also mention that $T_m^{'}$ increases with increasing Ni content for a certain fixed frequency and a similar behavior has been found for the $T_m^{''}$ dependence, as seen in the plots of Fig. 3. Therefore, both peaks in $\chi_{ac}$ data shift to higher temperatures with $x$, signaling the effect of interparticle (dipolar) interactions throughout the series.

As mentioned above, the observed frequency-dependent behavior in the $\chi_{ac}$ data may, in principle, be attributed to (i) the blocking process of superparamagnetism arising from isolated, noninteracting, and weak-interacting NPs (Refs. 15 and 16); (ii) intraparticle spin-glasslike behavior due to the reduced size of the NPs and surface disorder;[17] or (iii) collective spin-glass behavior, which is caused by strong dipolar interparticle interactions and randomness.[17] As few experimental methods are able to distinguish between the paramagnetic to spin-glass transition and the superparamagnetic to spin-glass transition, we further discuss our findings in the $\chi_{ac}$ data. Thus, from several curves similar to the ones displayed in Fig. 2, we have computed pairs of ($T_m^{''}$, $f$) for the construction of ln ($\tau$) vs 1/$T_m^{''}$curves for the five samples studied, as shown in Fig. 3. We mention here that a similar behavior has been found when the temperature maximum $T_m^{''}$ is been replaced by $T_m^{'}$ in curves of Fig. 3. The attempt time $\tau_0^N$ [see Fig. 4(a)] and the effective magnetic anisotropy $K_{\text{eff}}^N = (E_a^N/\langle V\rangle)$ obtained from the linear fittings by using the Néel-Arrhenius law of Eq. (1) are listed in Table II. They show that increasing $x$ results in a decrease in $\tau_0^N$, as displayed in Fig. 4(a), and a smooth increase in $K_{\text{eff}}^N$ (see Table II), indicating a progressive strengthening of the interaction between granules with $x$.

The increase in the dipolar interactions with $x$ manifests itself in a clear deviation of the linear fit in the curves of samples with 7.9 and 12.8 wt % Ni at very high frequencies, in a similar fashion as observed in systems with high content of maghemite and Fe NPs.[7,16] The values of $K_{\text{eff}}^N$ for the three less concentrated samples, inferred from data shown in Table II, are similar (~ $5.5\times10^5$ erg/cm$^3$) and in excellent agreement with $K_{\text{eff}}$ ~ $4.5\times10^5$ erg/cm$^3$ expected for Ni NPs in Ref. 18. However, for the two more concentrated samples, the values of $K_{\text{eff}}^N$ are higher and reached ~ $12\times10^5$ erg/cm$^3$, for the most concentrated specimen. Such a value of $K_{\text{eff}}^N$ is much higher than the one of the bulk Ni, $K_1 = -8\times10^5$ erg/cm$^3$ (Ref. 12) but may be attributed, in principle, to the magnetic contribution of a disordered spin surface layer of the Ni NPs. On the other hand, the attempted time $\tau_0^N = 5.8\times10^{-15}$ s for the most concentrated sample 12.8 wt % Ni has an unphysical small value, further indicating that a simple Néel-Arrhenius law is not adequate to describe the dynamical properties in samples with higher Ni content. In addition, when the interparticle interaction is nonnegligible, a collective state may occur provided that the system has a large distribution of particle size and randomly distributed magnetic NPs.[2]

Increasing concentration of the magnetic material may lead to a collective behavior of the magnetic NPs. Such a process, that differs from the individual blocked ones, can result in a thermodynamic phase transition, as frequently discussed in spin-glass systems.[2,6,9,16] In order to test the hypothesis that all samples studied here can be classified as SPM systems, a criterion, which is model independent, can be used.[10] It relates the relative shift of the temperature of the maximum in $\chi''$, $T''_{\mathrm{m}}$, with the measured frequency $f$ as

$$\Phi = \frac{\Delta T''_{\mathrm{m}}}{T''_{\mathrm{m}} \Delta \log(f)} \tag{4}$$

where $\Delta T''_{\mathrm{m}}$ is the relative variance of $T''_{\mathrm{m}}$ per decade frequency $\Delta \log(f)$. In spin-glass systems the variation in the freezing temperature $T_f \sim T''_{\mathrm{m}}$ with $\tau$ is believed to be very small and therefore $\Phi < 0.05$. On the other hand, SPM systems exhibit large values of $\Phi$ between 0.10 and 0.13 for noninteracting NPs, and 0.05 and 0.13 for interacting ones.[10] The values of $\Phi$ obtained in our Ni NPs, evaluated from the $\chi_{\mathrm{ac}}$ data taken near 50 Hz, are displayed in Table II. The results indicate that $\Phi$ varies significantly with increasing $x$ being 0.13 for the most diluted specimen (1.9 wt % Ni), decreases with increasing Ni content, and assumes a value of 0.08 for the most concentrated sample (12.8 wt % Ni). At first glance, the results strong indicate that changes in $T''_{\mathrm{m}}$ when $\tau$ is varied are much larger than those expected for spin-glass systems. However, to reject the hypothesis of a spin-glasslike state below $T_f \sim T''_{\mathrm{m}}$ in our samples, a spin glass model that predicts a scaling of the $T''_{\mathrm{m}}$ data by using a critical relaxation time $\tau$ must be considered.[19] Such a model predicts a slowing down characteristic of classical spin-glass systems with intermediate to strong interparticle interactions and the relationship between $\tau$, $T''_{\mathrm{m}}$, and $T_{\mathrm{g}}$ is given by

$$\tau = \tau_0 \left[ \frac{T_f - T_g}{T_g} \right]^{-\alpha}, \tag{5}$$

where $T_g$ is the glass transition temperature, $\tau_0$ is the attempt time, and $\alpha$ is the dynamical exponent that varies between 4 and 12 for different spin glasses.[17] This critical law has been successfully applied to test various NPs systems exhibiting spin-glasslike behavior.[2,16,20] Although the fitting of Eq. (5) to our $\chi_{ac}$ data resulted in reasonable values for both $T_g$ and $\alpha$ (listed in Table II), the ones of the pre-exponential $\tau_0$, displayed in Fig. 4(b) and roughly varying fom $4.0\times10^{-4}$ to $1.5\times10^{-1}$ s, are orders of magnitude higher than expected ($10^{-10}$–$10^{-14}$ s),[21] eliminating the hypothesis of a spin-glasslike transition in our samples. Besides, the fitted parameters $\alpha$ and $\tau_0$ appear to be independent of $x$, as inferred from the data shown in Fig. 4(b).

Returning to the results displayed in Fig. 3 and listed Table II, one is also leading to conclude that (i) the Néel-Arrhenius law, given by Eq. (1), seems to be inappropriate to describe the results for the more concentrated samples and (ii) that the scenario of classical spin-glasses is improper to describe the dynamic properties of this series. The obtained values of the parameters $\tau_0^N$, as displayed in Fig. 4(a), and the effective anisotropy constant $K_{\mathrm{eff}}^N$ (see Table II) are essentially Ni content independent in samples with $x$=1.9, 2.7, and 4.0 wt % Ni. On the other hand, these parameters decrease/increase abruptly for the samples with 7.9 and 12.8 wt % Ni, or, in other words, increasing the magnetic interaction results in a clear break down of the applicability of the Néel-Arrhenius law.

To take into account the magnetic interaction between NPs, we have used the Vogel-Fulcher law [Eq. (3)] to fit the same set of the ln ($\tau$) vs $1/T_m''$ curves, which are also displayed in the Fig. 3. The resulting fit parameters $\tau_0^V, K_{\mathrm{eff}}^V$, and $T_0^V$ are displayed in Table III. The excellent agreement between the fitting result and the experimental data indicates that considering the dipolar interaction between granules is necessary and adequate for describing the dynamical properties of our samples. The very small values of $T_0^V\sim$ 1 K in samples with 1.9, 2.7, and 4.0 wt % Ni further suggest that the strength of the dipolar interaction is weak in the limit of low Ni concentration (below $x\sim$ 4 wt % Ni). This result is expected since the dynamical properties of the NPs were well described by the simple Néel-Arrhenius law (see Table II), indicating that systems in the limit of very weak or negligible dipolar interactions may be described by both laws. Analyzing the data displayed in Table III and Fig. 4(a), we notice that $\tau_0^V$ is now independent of $x$ for all the series and assumes values within the expected range $10^{-12}$ –$10^{-9}$ s. The difference between the dependence of $\tau_0^N$ and $\tau_0^V$ with $x$, obtained from the Néel-Arrhenius and Vogel-Fulcher laws, are clearly seen in Fig. 4(a). A similar feature is observed in the effective anisotropy $K_{\mathrm{eff}}^{V}$ (see Table III), in which values are Ni content $x$ independent, and assumes essentially the same magnitude of the cubic anisotropy constant of bulk Ni ($K_1$= - 8×10$^5$ erg/cm$^3$).

As displayed in Table III and Fig. 5, similar results for the strength of the dipolar interaction were obtained from the fittings using the Vogel-Fulcher law ($T_0^V$) and Eq. (2) ($T_0^{Eq}$). A monotonically increase in $T_0^{Eq}$ with increasing Ni content was obtained and $T_0^V$ seems to be Ni concentration independent, at least for $x$ < 4wt % Ni. For higher Ni concentrations, $T_0^V$ further increases with increasing Ni concentration. Based on these results, one can define a kind of threshold limit value for the Ni concentration in which the dipolar interactions can be considered negligibly small. Such a limit was found here to be ~ 4 wt % Ni, in excellent agreement with others found in different systems where, at sufficiently low volume fraction ($x_V$ < 1%),[22] the NPs assemblies exhibit ideal SPM behavior and the relaxation is only governed by the thermally activated dynamics of the individual NPs.[6,20] In addition to this, the strength of the dipolar interactions in samples with $x$=7.9 and 12.8 wt % Ni seems to be not strong enough to lead to a collective spin-glass state. Although all the samples studied here have very low Ni concentrations, in the range of 0.5≲ $x_V$ ≲3.5% by volume, the dipolar interaction becomes important to be considered in samples with concentrations higher than 4 wt % Ni (or $x_V$ > 1%). Such a value is frequently found in other magnetic systems considered in the limit of weak interaction.[2,7,16]

It is important to remark that although the Ni concentration of 4 wt % Ni be a value in which the dipolar interaction can be considered negligibly small, we have seen that the dipolar interparticle interaction is actually present and systematically observed in the magnetic properties throughout the series. We finally mention that the magnetic dynamics in very diluted Ni NPs discussed here may be of interest for other systems, ranging from artificial spin ice[23] to particulate magnetic carriers in biological tissues,[24] where the occurrence and strengthening of the dipolar interactions are important.

## IV. CONCLUSION

From the results and discussions made above it is possible to conclude that dipolar interactions throughout nanoparticles modify their magnetic properties. The net result of the increasing dipolar interaction is mirrored in several features of the magnetic system as: (i) unphysical values of the attempting time $\tau_0$; (ii) an increase in both $T'_m$ and $T''_m$ (for a fixed *f*); (iii) a decrease in the relative variance of both both $T'_m$ and $T''_m$ per decade frequency; (iv) an increase in the anisotropy energy constant as a consequence of an apparent increasing energy barrier; and (v) a clear deviation from linearity in classical Néel-Arrhenius plots of ln ($\tau$) vs $1/T''_m$).We have also found that, in the limit of weak dipolar interactions, the Vogel-Fulcher law has been successfully applied to account for the dipolar interactions of an ensemble of Ni NPs with random uniaxial anisotropy and particles volume distribution. From the ac magnetic susceptibility data an upper limit value of ~ 4 wt % Ni has been estimated to be the concentration where dipolar interactions can be neglected. On the other hand, we have also noticed that the effect of increasing dipolar interaction may be seen throughout the series even below this upper limit. Although the characteristics of randomness and polydispersity presented in our samples, features of glassy behavior were not observed probably due to the weak interparticle dipolar interactions between granules.

## ACKNOWLEDGMENTS

The authors acknowledge support from the Brazilian agencies Fundação de Amparo à Pesquisa do Estado São Paulo (FAPESP) under Grant No. 2005/53241-9, Conselho Nacional de Desenvolvimento Científico e Tecnológico (CNPq) under Grant No. 473932/2007-5, and Coordenação de Aperfeiçoamento de Pessoal de Nível Superior (CAPES). One of us (R.F.J) was supported by CNPq under Grant No. 308706/2007-2.

## List of Figures

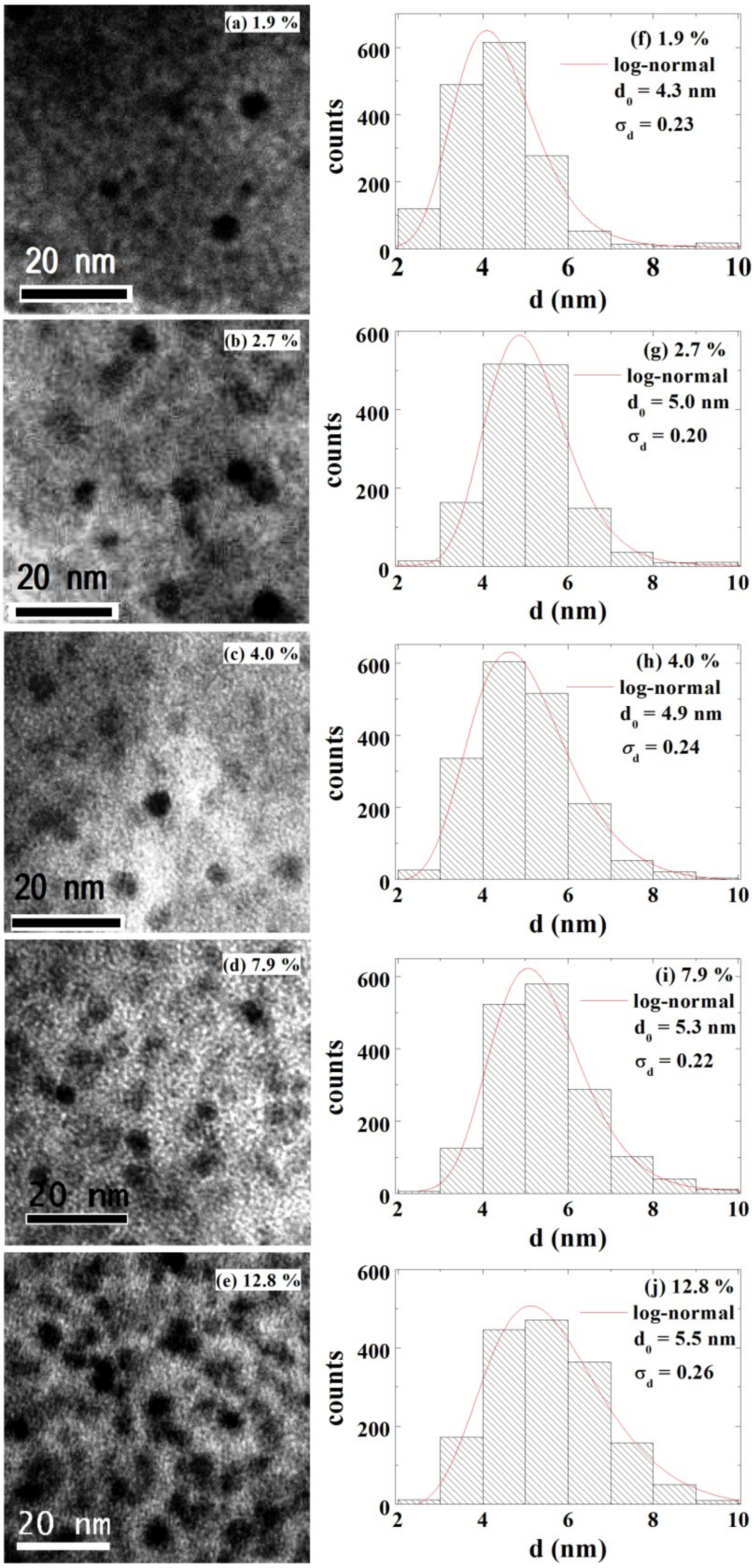

FIG. 1. [(a) – (e)] TEM images of the samples with 1.9, 2.7, 4.0, 7.9, and 12.8 wt % Ni. The images show nearly spherical Ni NPs with uniform sizes. [(f) – (j)] Histograms of the NP size

distributions for the correspondent samples determined from TEM images. Solid lines are the fittings by using a log-normal distribution.

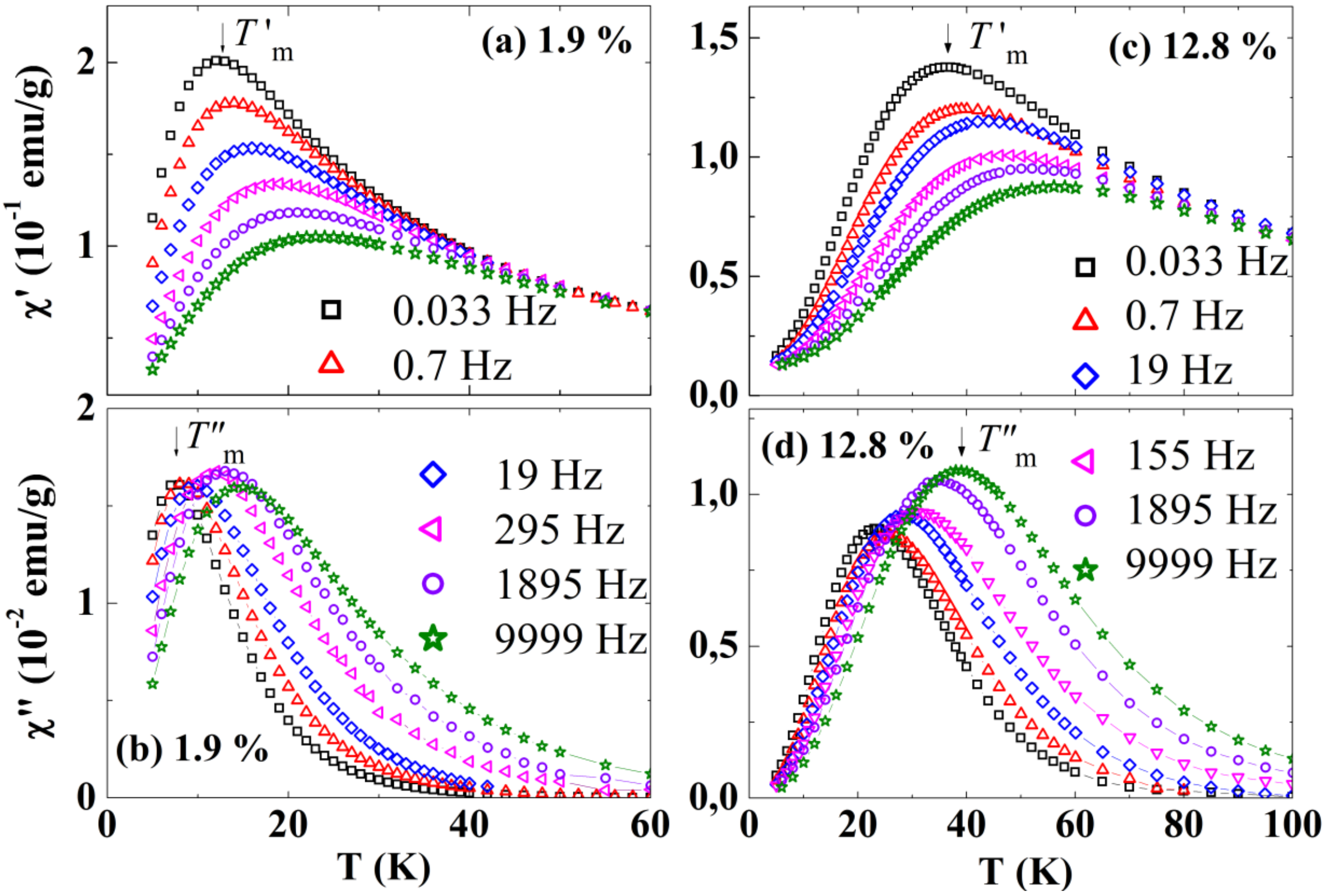

FIG. 2. Temperature dependence of the ac magnetic susceptibility for samples with 1.9 and 12.8 wt % Ni. Real (a) and (c) and imaginary (b) and (d) parts of the ac magnetic susceptibility at some selected frequencies are indicated in the figure.

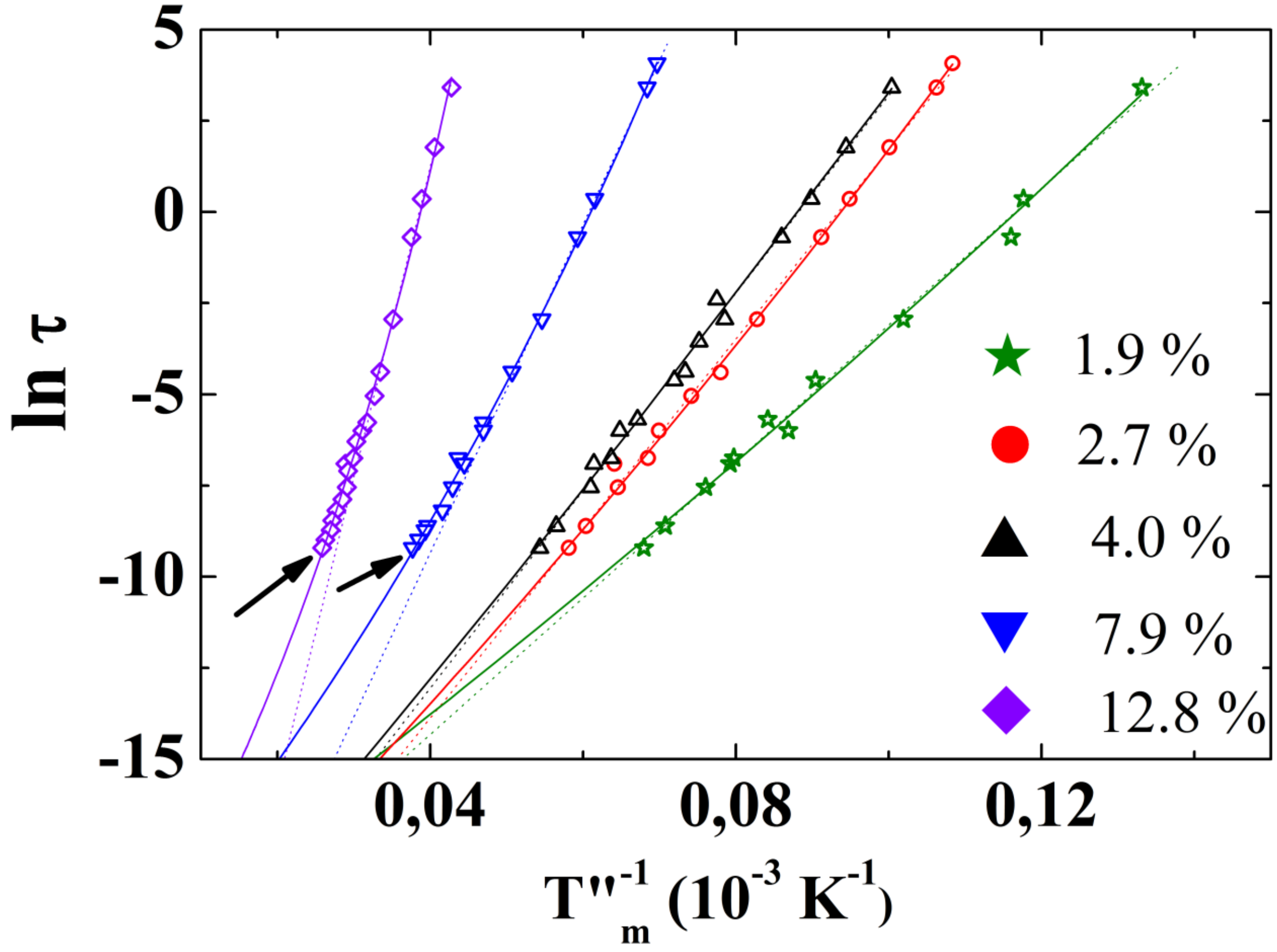

FIG. 3. Classical plots of ln ($\tau$) vs $1/T''_{\mathrm{m}}$ for samples with 1.9, 2.7, 4.0, 7.9, and 12.8 wt % Ni. Dotted lines are fitted to the Néel-Arrhenius law given by Eq. (1) and arrows indicate a clear deviation of the fitting in the limit of high frequencies. Solid lines represent the best fit of the Vogel-Fulcher law given by Eq.(3).

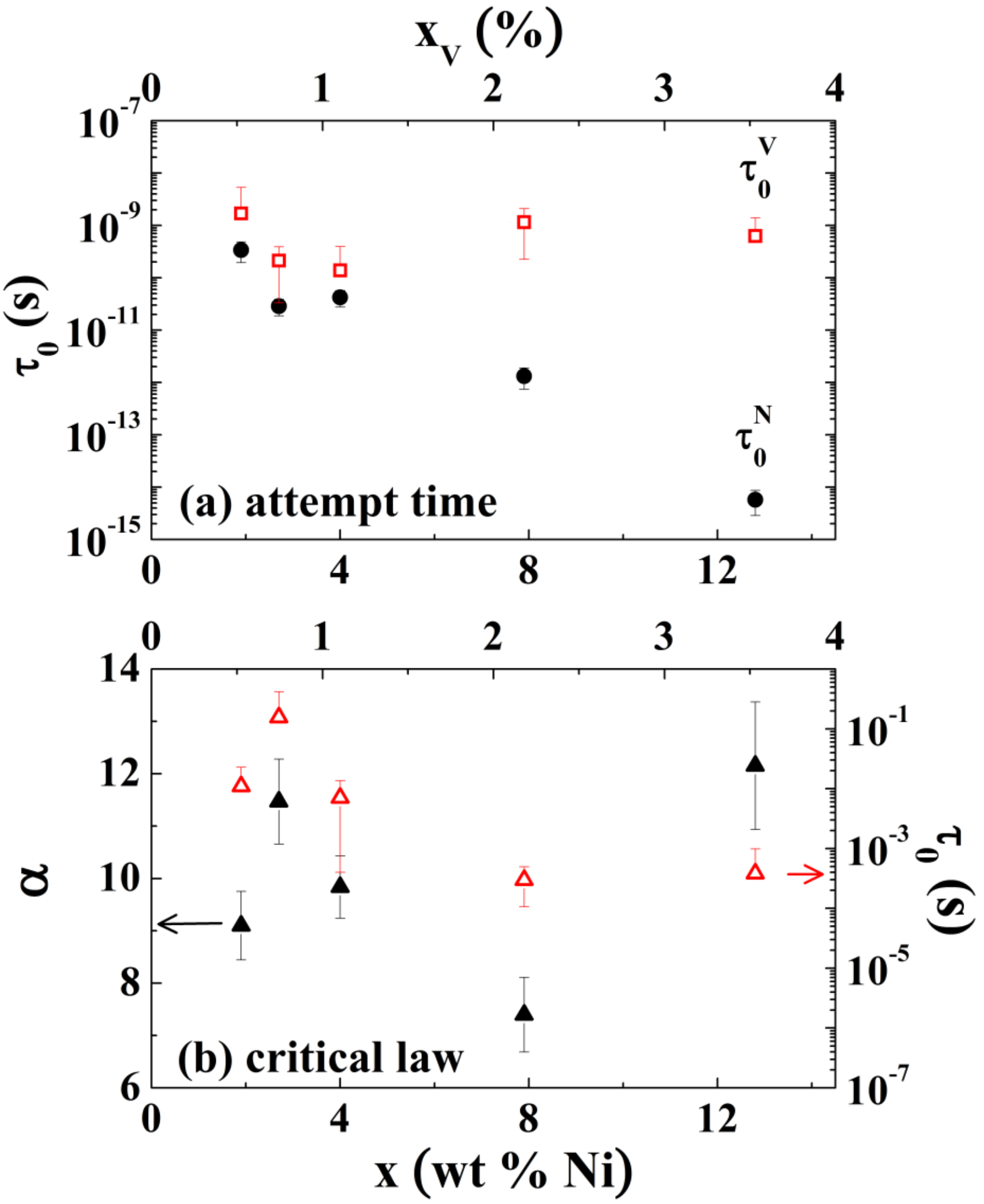

FIG. 4. (a) Attempt time obtained by fitting the data displayed in Fig. 3 to the Néel-Arrhenius law, $\tau_0^N$ (closed circles), and to the Vogel-Fulcher law, $\tau_0^V$ (open squares) as a function of the Ni NP concentration. (b) Dynamical exponent α (closed triangles) and attempt time $\tau_0$ (open triangles) as a function of the Ni NP concentration, obtained from the fittings of the critical law given by Eq. (5). The volume concentration $x_V$ as well as the weight concentration $x$ dependence of all parameters are displayed in the bottom and top axes, respectively. The error bars are also indicated.

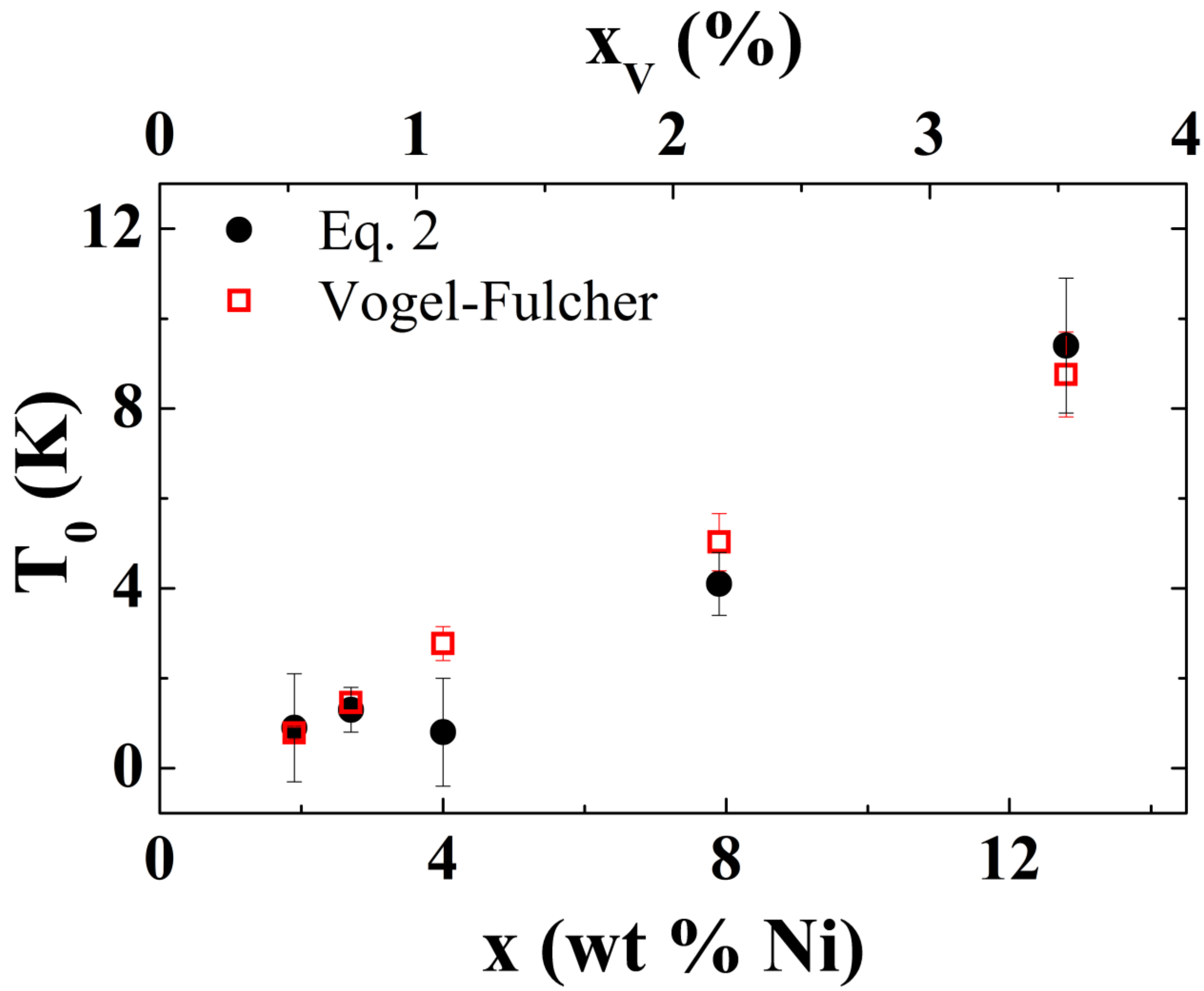

FIG. 5. Strength of the dipolar interaction $T_0$ as a function of the Ni NP concentration. Values of $T_0$ were obtained from both Eq. (2) ($T_0^{Eq}$, closed circles) and from the Vogel-Fulcher law ($T_0^V$, open squares). The volume concentration $x_V$ as well as the weight concentration $x$ dependence of $T_0$ are displayed in the figure. The error bars are also indicated.

**List of Tables**

TABLE I. Median NP diameter $d_0$, width $\sigma_d$, mean value $\langle d \rangle$, and average interparticle distance $r$ extracted from TEM analysis.

| ICP-AES | | TEM | | |
|---|---|---|---|---|
| $x$ (wt % Ni) | $r$ (nm) | $d_0$ (nm) | $\sigma_d$ | $\langle d \rangle$ (nm) |
| 1.9 | 21 | 4.3 | 0.23 | 4.4 |
| 2.7 | 21 | 5.0 | 0.20 | 5.1 |
| 4.0 | 18 | 4.9 | 0.24 | 5.0 |
| 7.9 | 16 | 5.3 | 0.22 | 5.4 |
| 12.8 | 14 | 5.5 | 0.26 | 5.7 |

TABLE II. Values of  evaluated at $f \sim 50$ Hz accompanied by the fitted parameters of both the Néel-Arrhenius and the critical law given by Eq. (5) for the series studied.

| | | Néel-Arrhenius | | Critical law | | |
|---|---|---|---|---|---|---|
| $x$ (wt % Ni) | Φ | $\tau_0^N$ (s) | $K_{\rm eff}^N$ ($-10^5$ erg/cm$^3$) | $\tau_0$ (s) | $T_g$ (K) | α |
| 1.9 | 0.13 | $3.4\times10^{-10}$ | 5.9 ±1.2 | $1.1\times10^{-2}$ | 8.7 | 9.1 |
| 2.7 | 0.11 | $2.9\times10^{-11}$ | 5.2 ±0.9 | $1.5\times10^{-1}$ | 9.1 | 11.5 |
| 4.0 | 0.12 | $4.2\times10^{-11}$ | 5.7 ±1.0 | $7.0\times10^{-2}$ | 11.2 | 9.8 |
| 7.9 | 0.10 | $1.1\times10^{-11}$ | 7.5 ±1.3 | $3.8\times10^{-4}$ | 19.3 | 7.4 |
| 12.8 | 0.08 | $2.1\times10^{-13}$ | 12.1 ±1.9 | $4.0\times10^{-4}$ | 26.2 | 12.1 |

TABLE III. Attempt time $\tau_0^V$, effective anisotropy constant $K_{\text{eff}}^V = E_a/V$, and $T_0$ extracted from the Vogel-Fulcher law ($T_0^V$). Another estimated values of $T_0^{Eq}$, obtained by using Eq. (2) with $M_S(0)$ = 509 emu/cm$^3$ of bulk Ni, are also displayed.

| | Vogel-Fulcher | | | Eq. (2) |
|---|---|---|---|---|
| $x$ (wt % Ni) | $\tau_0^V$ (s) | $K_{\text{eff}}^V$ (-10$^5$ erg/cm$^3$) | $T_0^V$ (K) | $T_0^{Eq}$ (K) |
| 1.9 | $1.7 \times 10^{-9}$ | 4.8 ± 1.2 | 0.9 | 0.8 |
| 2.7 | $2.1 \times 10^{-10}$ | 4.1 ± 0.4 | 1.3 | 1.5 |
| 4.0 | $1.4 \times 10^{-10}$ | 5.1 ± 1.0 | 0.8 | 2.8 |
| 7.9 | $1.2 \times 10^{-9}$ | 4.2 ± 0.4 | 4.1 | 5.0 |
| 12.8 | $6.3 \times 10^{-10}$ | 4.9 ± 0.7 | 9.4 | 8.8 |